\documentclass[12pt]{JHEP3}
\usepackage{epsfig}
\usepackage{amssymb}
\usepackage{bm}
\usepackage{amsmath}
\newcommand{\p}{\partial}
\def\beq{\begin{eqnarray}}
\def\eeq{\end{eqnarray}}

\makeatletter
\@addtoreset{equation}{section}
\makeatother

\newcommand{\wh}{\widehat}

\newcommand{\del}{\partial}

\newcommand{\half}{\frac{1}{2}}

\def\tr{\mathop{\rm tr}\nolimits}

\newcommand{\cF}{{\mathcal F}}

\newcommand{\cA}{{\mathcal A}}

\newcommand{\Mkk}{M_{\rm KK}}

\def\matt[#1,#2,#3,#4]{\left(%
\begin{array}{cc} #1 & #2 \\ #3 & #4 \end{array} \right)}

\title{
Quark mass dependence of hadron spectrum in holographic QCD
}
\author{
Koji Hashimoto$^{*,a}$, 
Takayuki Hirayama$^{\dagger,b}$ and 
Deog Ki Hong$^{\S,c}$\\
${^*}$ {\it Theoretical Physics Laboratory, RIKEN, 
Saitama 351-0198, Japan}\\
${}^\dagger$ 
{\it Physics Division, National Center for Theoretical Sciences, }
\\ \hspace{80mm}{\it Hsinchu 300, Taiwan}\\ 
${}^\dagger$
{\it Department of Physics, National Taiwan Normal University, }
\\ \hspace{80mm}{\it Taipei 116, Taiwan}\\
${}^\ddagger$ {\it Department of Physics, Pusan National University,
Busan 609-735, Korea}\\
$^a$ E-mail: \email{koji@riken.jp}\\
$^b$ E-mail: \email{hirayama@phys.cts.nthu.edu.tw}\\
$^c$ E-mail: \email{dkhong@pusan.ac.kr}\\
}

\abstract{
We compute a shift of baryon mass spectra due to quark masses in perturbation, 
in Sakai-Sugimoto model of holographic QCD. We find the shift for the
ground state nucleons to be  
$\delta M=4.1~m_{\pi}^2\,{\rm GeV}^{-1}$, which 
is consistent 
with the current lattice QCD result.
We predict the same value of the shift for N(1535) and $\Delta$,
while a larger value
$7.7~m_{\pi}^2\,{\rm GeV}^{-1}$ for Roper N(1440). 
We also present some evidences that the shifts of the vector meson
masses are suppressed in the large 't Hooft coupling limit. 
}

\preprint{
{\normalsize PNUTP-09/A04} \\
{\normalsize RIKEN-TH-157}\\
{\normalsize }
}

\begin{document}


\section{Introduction}

Recent progress in holographic QCD of string theory enables us to
compute the 
important physical observables of QCD: the hadron masses. 
QCD has only two kinds of dimension-ful parameters: the prime one is the
QCD scale which basically generates all the hadron masses, and the quark
masses. For the light quarks, the quark masses can be treated as a
perturbation (while heavy quarks have their particular importance in
various phases in QCD). In this short paper, we present a computation of
the hadron mass shift due to the quark masses, in the Sakai-Sugimoto
model \cite{SaSu1,SaSu2} which is the most successful model of
holographic QCD so far.  

Sakai-Sugimoto model is a holographic dual of {\it massless} QCD. 
There proposed two ways to introduce the quark mass to the model: (i)
worldsheet instantons \cite{Aharony:2008an,Hashimoto:2008sr} (see also
\cite{McNees:2008km,Argyres:2008sw}),  
and (ii) tachyon condensation of the
D8/$\overline{\rm D8}$-branes \cite{Casero:2007ae}. We are going to use
the 
worldsheet instanton method (i), 
since the other method (ii) assumes 
a
tachyon ``effective'' action which
is difficult to be validated in any manner in string theory except for
using string field theories.\footnote{The tachyon approach has more
direct analogue with a famous bottom-up model of holographic QCD
\cite{Erlich:2005qh}, and the mechanism is similar to the Higgs
mechanism, while the worldsheet instanton approach mimics extended
technicolor, see \cite{Hashimoto:2008sr, Hirayama:2007hz}.}

The worldsheet instanton produces the quark mass term in QCD, and the
gravity dual of the worldsheet instanton generates a term of the form
\begin{eqnarray}
 \delta L = c \, {\rm tr}\left[M U\right]
 \label{cmu}
\end{eqnarray}
where $M$ is the quark mass matrix of the form $M=$diag$(m_{\rm u},
m_{\rm d})$, and $U$ is the pion field in the standard notation in the
chiral perturbation theory: $U\equiv \exp [ 2i \pi(x)/f_\pi]$.
The value $c$ is computed from the worldsheet instanton amplitude,
but it turned out to be
difficult to evaluate (for some computations, see 
\cite{Hashimoto:2008sr,Aharony:2008an,McNees:2008km}).
Except for the pion mass through~\eqref{cmu}, the contribution of quark
masses to other meson masses has 
not been
computed yet.

In this paper, we present two computations. 
First, in Sec.~\ref{sec:2}, we 
compute shift of the baryon mass spectra
due to the quark mass, by assuming that 
the quark mass dependence in the meson effective lagrangian of
the Sakai-Sugimoto model appears only in the new term (\ref{cmu}),
at the leading order in expansion in $1/\lambda$. 
This assumption 
will be discussed in 
Sec.~\ref{sec:3}. There we will provide some computations 
and
show that shift of the vector/axial vector meson masses is small and
at a higher order in $1/\lambda$ expansion. 

Under the assumption mentioned above, it is quite easy to compute the
baryon mass shift. As noted in \cite{SaSu1} and computed in detail in
\cite{HSSY},\footnote{See also \cite{Hong:2007kx,Hong:2007ay} 
for an alternative
description of baryons by
introducing 5-dimensional fermion fields.
Other work for introducing baryons in holographic QCD includes
\cite{Hong:2006ta}. } 
the baryon in the Sakai-Sugimoto model is nothing but an
instanton-like soliton in the 5-dimensional Yang-Mills-Chern-Simons 
(YMCS)
action describing all the mesons effectively in a unified way.
The term (\ref{cmu}) can be thought of as a perturbation to the
YMCS action, so the mass shift can be computed by just inserting the
soliton configuration obtained in \cite{HSSY} to (\ref{cmu}).
The resultant correction is evaluated with baryon wave functions of
\cite{HSSY}.
Since the parameter $c$ is undetermined, 
we
obtain
the baryon mass as a function of the pion mass. Our result for
nucleons is in rough
agreement with results of lattice QCD.  
The mass shifts for N(1535) and $\Delta$ turn out to be the same as
that of the nucleons, while the shift for Roper N(1440) is larger.

\section{Baryon Mass Shift}
\label{sec:2}

\subsection{Set-up in Sakai-Sugimoto model}
\label{sec:2-1-1}

\subsubsection{A review of baryons}

The Sakai-Sugimoto model \cite{SaSu1,SaSu2} is described by 
the following five-dimensional $U(N_f)$ YMCS action in a curved
background,
\begin{align}
S=S_{\rm YM}+S_{\rm CS} \ , \quad
S_{\rm YM}=-\kappa
\int d^4 x dz\,\tr\left[\,
\half\,h(z){\cF}_{\mu\nu}^2+k(z){\cF}_{\mu z}^2
\right]\ .
\label{model}
\end{align}
$S_{\rm CS}$ is 
the 5-form Chern-Simons term $S_{\rm CS}$.
The $U(N_f)$ gauge fields $\cA_\mu (\mu=0,1,2,3)$ and $\cA_z$
have their field strength defined in the standard manner, $\cF_{\mu\nu}
= \partial_\mu \cA_\nu - \partial_\nu \cA_\mu - i[\cA_\mu,\cA_\nu]$.
The extra dimension $z$ unifies all the vector/axial vector mesons via
Kaluza-Klein like decomposition of the gauge fields, through the
nontrivial functions $h(z)=(1+z^2)^{-1/3}$ and $k(z)=1+z^2$.
The constant $\kappa$ is defined as 
$\kappa=a \lambda N_c$ with $a=1/(216\pi^3)$,
where $\lambda$ is the 't~Hooft coupling constant.
The dimensionful parameter $M_{\rm KK}$ in the action (\ref{model})
is already put to be the unity, and it was argued in \cite{SaSu1,SaSu2}
that the values $\Mkk=949\mbox{ MeV}$ and $\kappa=0.00745$
can fit the observed values of the $\rho$ meson mass 
$m_\rho\simeq 776 {\rm MeV}$ and the pion decay constant $f_\pi\simeq
92.6~{\rm MeV}$. 
In this paper we consider $N_f=2$, the two-flavor case,
and then
the $U(2)$ hermitian matrix $\cA$ is decomposed by Pauli matrices 
$\tau^a$ ($a=1,2,3$) and a unit matrix ${\bf 1}_2\equiv \tau^0$,
\begin{eqnarray}
\cA=A+\wh A\,\frac{{\bf 1}_2}{2}
=A^a\frac{\tau^a}{2}+\wh A\,\frac{{\bf 1}_2}{2} \ .
\label{decom}
\end{eqnarray}

The baryon is identified as a soliton solution localized in the
4-dimensional $x^M$ space ($M=1,2,3,z$) \cite{SaSu1}. 
The instanton number of the Yang-Mills theory (\ref{model})
is
identified with the baryon number. 

The authors of \cite{HSSY} found an explicit solution
of the equations of motion of (\ref{model}).
In view of the fact that the size of the baryon scales as
$\lambda^{-1/2}$ as found in \cite{Hong:2007kx,HSSY}, 
an appropriate rescaling of the gauge fields and the coordinate by a
factor of $\lambda$ was identified in \cite{HSSY}. But here, for our
later purpose, we write the un-rescaled solution.
The solution is almost identical to a BPST instanton except for
a $1/\lambda$ ``correction'' appearing in the $U(1)$ part of the zeroth
component of the gauge field. 
This sub-leading contribution can be
understood from the fact that the solution represents a baryon whose
charge should be identified with the electric charge of the trace 
of the vector part of the chiral symmetry. 
The explicit solution is
\begin{align}
A_M^{\rm cl}=&-if(\xi)g\del_M g^{-1} 
, \quad
\wh A_0^{\rm cl}=
\frac{1}{\lambda}\frac{1}{8\pi^2 a}
\frac{1}{\xi^2}
\left[1-\frac{\rho^4}{(\rho^2+\xi^2)^2}
\right]
,~~~
A_0=\wh A_M=0 
,
\label{HSSYsol}
\end{align}
and the
non-Abelian part is 
given by
the BPST instanton, 
\begin{align}
f(\xi)=\frac{\xi^2}{\xi^2+\rho^2}\ , \quad
g(x)=\frac{(z-Z){\bf 1}_2+i(\vec{x}-\vec{X})\cdot\vec\tau}{\xi} \ .
\label{defg}
\end{align}
Here the moduli parameters of the instanton come in explicitly: 
the instanton size $\rho$ and the instanton location 
$X^M=(X^1,X^2,X^3,Z)=(\vec X,Z)$. $\xi$ is the distance in this
4-dimensional space, $\xi=
\sqrt{(z-Z)^2+|\vec{x}-\vec{X}|^2}$.
In addition to these 5 moduli
parameters, the rotation in the gauge $SU(2)$ space 
are
moduli,
which are incorporated as $V\in SU(2)$, 
\begin{eqnarray}
A_M=VA_M^{\rm cl}V^{-1}-iV\del_M V^{-1} \ .
\label{HSSY:AM}
\end{eqnarray}
Using the standard technique for quantizing solitons \cite{GeSa,Manton}, 
quantized baryon states appear in the degrees of freedom of
$(Z,\rho,V)$, a part of which acquires 
definite potentials in their quantum mechanics
via the effect of the curved spacetime and $S_{\rm CS}$ \cite{HSSY}. 
For example, 
the classical size and the location $Z$ of the instanton can be fixed by
minimizing the potential as
\begin{eqnarray}
\rho_{\rm cl}^2
=\frac{1}{\lambda}\frac{1}{8\pi^2 a}\sqrt{\frac{6}{5}}\ , \quad
Z_{\rm cl}=0\ ,
\label{rhocl}
\end{eqnarray}
so the size of the baryon is ${\cal O}(\lambda^{-1/2})$.
The spectrum of the baryon can be computed by evaluating
the hamiltonian of $(Z,\rho,V)$ with their quantum wave functions.

\subsubsection{A review of quark mass}
\label{sec:2-1-2}

We introduce the quark mass to the Sakai-Sugimoto model by the
worldsheet instantons. The method was first introduced in 
\cite{Aharony:2008an} and \cite{Hashimoto:2008sr}. In these papers, 
there are some differences in regularization of the worldsheet
instantons, and in this paper we follow \cite{Hashimoto:2008sr}.

The quark mass can be introduced by joining the D8-brane and the
$\overline{\rm D8}$-brane by
D6-branes
\cite{Hashimoto:2008sr}.
Once the D8 and the $\overline{\rm D8}$ are joined in this way, it is
possible to put a Euclidean string worldsheet whose boundary is defined
by the color D4-branes, the D8-brane, the D6-brane and the
$\overline{\rm D8}$-brane: these four sets of D-branes form a
square, 
in the middle of which one can put the
worldsheet instanton. Since the left-handed and right-handed quarks live
at the D8-D4 and $\overline{\rm D8}$-D4 intersections 
respectively, 
the worldsheet
instanton involves with $\bar{q}_L q_R$ vertex which is nothing but the
quark mass operator. The instanton amplitude, which is proportional to
$\exp[-A/2\pi\alpha']$ where $A$ is the area of the square, gives the
quark mass. Therefore the quark mass can be chosen by tuning the
location of the D6-branes
and their number. 

In the gravity dual, the introduced D6-brane is still a probe D-brane,
so one can put the worldsheet instanton in the same manner,
but now two of
the corners
are smeared out by the background curved geometry.
The meson excitation described by the gauge fields on the D8-branes
induces a worldsheet boundary coupling in the worldsheet instanton
amplitude, 
\begin{eqnarray}
 \delta S = c \int d^4x \;
{\rm Ptr} 
\left[M
\left(
\exp\left[-i \int_{-z_m}^{z_m}\cA_z dz\right]
-{\bf 1}_2\right)
\right] + \mbox{c.c.}, \;\;
M \equiv \left(
\begin{array}{cc}
m_u & 0 \\ 0 & m_d
\end{array}
\right)
\label{cma}
\end{eqnarray}
where $m_u$ and $m_d$ are quark masses, and $z=\pm z_m$ 
is the location of the D6-brane ($z=z_m$ and $z=-z_m$ specify the same
radial position, on the D8 and the $\overline{\rm D8}$).
This exponential form is a familiar boundary coupling of string
worldsheet, and the integral is over the period where the string
worldsheet ends on the D8-brane. For 
nonzero values of the
quark mass, the chiral symmetry
is explicitly broken, which is
consistent with the standard expectation for the quark mass term in
QCD. In \cite{Hashimoto:2008sr} a rough evaluation of the constant $c$
was given. The 
constant
subtraction $-{\bf 1}_2$ is for our convenience, to make
sure that when $\cA_z=0$ the instanton amplitude vanishes.

The $\cA_z$ field in nothing but the pion field \cite{SaSu1,SaSu2}, and 
the relation is 
\begin{eqnarray}
 {\rm P}\exp\left[
-i\int_{-\infty}^\infty dz \cA_z
\right] = \exp\left[
2i \pi(x)/f_\pi
\right] \equiv U \ ,
\label{defU}
\end{eqnarray}
where
$U$ is the standard notation for the pion field in the chiral
perturbation theory. Using this, the additional action to the
Sakai-Sugimoto model (\ref{model}) is
\begin{eqnarray}
 \delta S = \int\!  d^4x \; \delta L \ , 
\quad
\delta L \equiv c 
 \; 
{\rm tr}\left[M(U + U^\dagger-2 {\bf 1}_2)\right] \ ,
\label{mu}
\end{eqnarray}
which is the well-known term in the chiral perturbation theory.
Here although $z_m$ is a finite value, we approximate the exponential
part in~\eqref{cma} by~\eqref{defU}. For the purpose of a rough estimate,
this simplification can be justified since the pion wavefunction is
localized at $z=0$. 

Using the pion mass $m_\pi$, we
obtain a relation
\begin{eqnarray}
 c (m_u + m_d)= \frac12 f_\pi^2 m_\pi^2.
\label{qmass}
\end{eqnarray}
We are working in the normalization 
$\pi(x)= \pi^a(x) T_a$, ${\rm tr}[T_aT_b]= \frac12  \delta_{ab}$.

We have seen that the introduction of the D6-brane ending on the D8 and
the $\overline{\rm D8}$-branes can induce a quark mass term. However,
the additional action (\ref{cma}) should not be the only correction to
the Sakai-Sugimoto action. 
In fact there
are possibly two other sources for new
terms coming from the presence of the D6-brane:
\begin{itemize}
\item Vector meson / axial vector meson mass terms.
\item Shift of D8-brane configuration pulled by the worldsheet
       instanton. 
\end{itemize}
The first one is very interesting, as it may give an interesting result
on shift of the meson mass spectra due to the quark mass. However, in 
the next section, we are going to discuss that this shift is suppressed
by $1/\lambda$ in the holographic QCD. The second one, the shift of the
location (shape) of the D8-branes, is intriguing of its own and was
studied in \cite{Aharony:2008an}. Since this is obviously a secondary
effect in the sense that the string worldsheet instanton
pulls the heavy D8-branes, we will not consider it in this
paper.\footnote{
This tiny modification of the shape of the D8-brane can be numerically
computed, and in fact it increases the mass of baryon. See
\cite{Seki:2008mu} for a relevant discussion. This will also 
change the meson masses.
}

In the following in this section, we will ignore these two additional
possibilities and consider only (\ref{cma}) for computing the baryon
mass shift.

\subsection{Baryon mass shift : classical evaluation}
\label{sec:2-2}

When the quark mass is small, 
at the leading order in the quark
mass, the shift of the baryon mass is simply given by
\begin{eqnarray}
 \delta M = - \int d^3x \; \delta L[A^{\rm cl}]
\label{barc}
\end{eqnarray}
where $A^{\rm cl}$ is the classical solution (\ref{HSSYsol})
representing the baryon, and $\delta L$ is given by 
(\ref{cma}). 
By the additional term (\ref{cma}),
the baryon configuration 
itself would be modified
and it would shift the baryon mass at the higher order in the
quark mass, but we neglect the higher order terms in this paper.

If we substitute the classical solution to (\ref{cma}), the
resultant expression is a function of the instanton moduli. At the
classical level (which is the large $N_c$ limit), we can substitute the
classically fixed moduli 
(\ref{rhocl}). To distinguish different baryon states, we use
baryon wave functions of the moduli given in \cite{HSSY} and take the
expectation value of (\ref{cma}), which gives the baryon mass shift for
a given baryon state. Here we shall compute the classical value first,
and then in the next subsection we will compute the quantized value for
each baryon state. 

Let us evaluate this (\ref{barc}) explicitly. 
First, note that in the large $\lambda$ expansion, the leading order 
term of the solution is just the same as the BPST instanton, 
while the term of the next-to-leading order is 
nonzero only for the zero-th component $\widehat{A}_0(x,z)$ of 
the overall $U(1)$ part of the gauge field. 
On the other hand, the
integral in defining $U$ in (\ref{defU}) which appear in (\ref{barc}) 
is only for the $z$ component. Thus, the
sub-leading term does not contribute to our calculation, so 
what we need is just the BPST instanton part. 

Eventually, the integral (\ref{defU}) was already evaluated in
\cite{AM}, so let us use it. For simplicity, we put all the moduli to be
the classical
values, (\ref{rhocl}) and $V={\bf 1}_2$, to obtain a classical
shift of the baryon mass (we can put $X^a=0$ without losing generality). 
We work in a singular gauge of the
BPST instanton, 
\begin{eqnarray}
 A_z = \left(
\frac1{\xi^2}-\frac1{\xi^2 +\rho^2}\right)
(x^a-X^a) \tau^a \ .
\label{A_z}
\end{eqnarray}
The $A_z$ configuration (\ref{A_z}) is proportional to a matrix
$(x^a) \tau^a$ for any value of $z$, so the evaluation of the path
ordering in (\ref{barc}) reduces to an Abelian problem \cite{AM}. 
We obtain 
\begin{eqnarray}
 U = \exp\left[i f(r) \hat{x}^a \tau^a\right]
\end{eqnarray}
with 
\begin{eqnarray}
 f(r)= \pi \left[
1-\frac{1}{\sqrt{1+\rho^2/r^2}}
\right], \quad r\equiv \sqrt{(x^1)^2 + (x^2)^2 + (x^3)^2}.
\end{eqnarray}
With this expression, we obtain
\begin{eqnarray}
 U+U^\dagger = 2 \cos(f){\bf 1}_2.
\label{unity}
\end{eqnarray}
Therefore, using the relation (\ref{qmass}), we have 
\begin{eqnarray}
 \delta M &=& \int d^3x \; f_\pi^2 m_\pi^2 
(1-\cos(f))
\nonumber \\
& = & 
4\pi f_\pi^2 m_\pi^2 \rho^3 
\int d\tilde{r} \; \tilde{r}^2 
\left(
1-\cos\left[\pi \left(1-\frac{1}{\sqrt{1+ \tilde{r}^{-2}}}\right)\right]
\right)
\end{eqnarray}
where we made a change of a variable as $r = \rho \tilde{r}$.
The last integral is numerically evaluated as 
\begin{eqnarray}
\int d\tilde{r} \; \tilde{r}^2 
\left(
1-\cos\left[\pi \left(1-\frac{1}{\sqrt{1+ \tilde{r}^{-2}}}\right)\right]
\right)
=1.104.
\end{eqnarray}
Using this, we obtain a formula for the shift of the baryon mass
spectrum due to the quark mass, 
\begin{eqnarray}
\frac{\delta M}{m_\pi^2} =  4 \pi f_\pi^2 \rho^3 \times
 1.104 \ . 
\label{shiftc}
\end{eqnarray}

In the classical limit $N_c\rightarrow \infty$, $\rho$ is equal to
$\rho_{\rm cl}$ which is given in (\ref{rhocl}), so
\begin{eqnarray}
 \delta M = 1.104 \times 4\pi f_\pi^2 m_\pi^2 \rho_{\rm cl}^3
= 1.104 \times 4\pi f_\pi^2 m_\pi^2 
\left(\frac{27\pi \sqrt{6}}{\sqrt{5}\lambda}\right)^{3/2} \ .
\end{eqnarray}
 In this limit, the mass
shift (\ref{shiftc}) is independent of the baryon state, since the
dependence on the baryon state appears at the quantum level which is
sub-leading in the $1/N_c$ expansion. In the next subsection we compute the
$1/N_c$ corrections and the baryon state dependence. It was found
that $1/N_c$ 
corrections are rather large for various static properties of baryons
(such as charge radii) in \cite{Hashimoto:2008zw} .

Note that our baryon mass shift $\delta M$ is linear in $m_\pi^2$ as
seen in (\ref{shiftc}), 
to the lowest order in $m_\pi$. 
This is in agreement with the lattice QCD results and also with 
chiral perturbation theories. 

\subsection{Baryon mass shift : baryon state dependence}
\label{sec:2-3}

The classical evaluation of the baryon mass shift, (\ref{shiftc}), will
be corrected once we take into account the quantum states of the baryon. 
Each baryon state is specified by the quantum numbers 
$\{I(=J), I_3, n_\rho,n_Z\}$ where $I$ is the isospin, $J$ is the spin, 
$n_\rho$ and $n_Z$ are quantum numbers associated with the moduli 
parameters $\rho$ and $Z$ \cite{HSSY}. 
In general the mass shift depends on these quantum numbers.

The dependence can be evaluated by the standard perturbation in
quantum mechanics. The additional hamiltonian is (\ref{barc})
evaluated with all the moduli dependence in $A_{\rm cl}$.

First, let us consider $Z$-dependence. Since the $Z$-dependence comes in
as $z-Z$ in the solution (\ref{HSSYsol}) with (\ref{defg}), once we
perform the integration (\ref{defU}), $Z$ disappears. So there is no
additional hamiltonian for $Z$. 

Next, let us consider $V$. This $SU(2)$ rotation $V$ appears in the
solution as a gauge transformation of the solution, (\ref{HSSY:AM}),
the transformation of $U$ is given as $VUV^{-1}$, so 
$U+U^\dagger \rightarrow V(U+U^\dagger)V^{-1}$. Now, looking at
(\ref{unity}), we notice that in fact this $V$ dependence disappear,
because $U+U^\dagger$ is proportional to the unit matrix. 
So, there is no additional hamiltonian for $V$.

Finally, let us consider the moduli $\rho$. The expression
(\ref{shiftc}) has a factor $\rho^3$, so it is the additional
hamiltonian for the quantum mechanics for $\rho$.
In sum, the shift of the baryon mass is dependent only on the quantum
number $n_\rho$. It is not sensitive to spin/isospin and also the
quantum number $n_Z$. This means that, in particular, proton, neutron,
$N(1535)$ and delta excitation have the same mass shift since they
share the same value $n_\rho=0$, while that of 
the Roper excitation $N(1440)$ ($n_\rho=1$) is different. In particular, it is
interesting that, 
although generic values of $(m_u,m_d)$ seem to break the isospin
invariance, our result shows that the leading order mass shift of the
baryons is insensitive to this breaking.

We need to evaluate $\langle \rho^3 \rangle$ with the baryon 
wave function $R_{n_\rho}(\rho)$. The explicit expression for the wave function is
given in \cite{HSSY}. For the lowest $n_\rho=0$, it is
\begin{eqnarray}
 R_0(\rho) = {\cal N}_0\;
\rho^{-1+2\sqrt{1+N_c^2/5}}\exp 
\left[-\frac{M_0}{\sqrt{6}}\rho^2\right]
\end{eqnarray}
where $M_0\equiv 8\pi^2 \kappa$, and ${\cal N}_{n_\rho}$ is the
normalization factor. The wave function is normalized as
\begin{eqnarray}
 \int d\rho \; \rho^3 R_{n_\rho}(\rho)^2 = 1 \ ,
\end{eqnarray}
with the factor $\rho^3$ which 
is the Jacobian for the 4-dimensional spherical
coordinate system in the one-instanton moduli space.
The expectation value is evaluated as
\begin{eqnarray}
 \left\langle \rho^3 \right\rangle_{n_\rho=0}
 & =& 
\frac{\displaystyle\int_0^\infty \!\!\!\! 
d\rho \;  \rho^6 R_0(\rho)^2}
{\displaystyle\int_0^\infty \!\!\!\!  d\rho \;  \rho^3 R_0(\rho)^2}
= \left(\frac{\sqrt{6}}{2M_0}\right)^{\!\!3/2} \!\!
\frac{\displaystyle\int_0^\infty \!\!\!\! 
dt \;  t^{3/2+ 2\sqrt{1+N_c^2/5}} e^{-t}}
{\displaystyle\int_0^\infty \!\!\!\! 
dt \;  t^{2\sqrt{1+N_c^2/5}} e^{-t}}
\nonumber \\
&=&
\rho_{\rm cl}^3\left(\frac{\sqrt{5}}{2N_c}\right)^{\!\!3/2} \!\!
\frac{\Gamma(b+3/2)}
{\Gamma(b)}
\ ,
\label{ratio1}
\end{eqnarray}
where in the second equality we rescaled the variable as 
$\rho = (\sqrt{6}/(2M_0))^{1/2}\; t^{1/2}$, and we have defined
$b\equiv 1+2\sqrt{1+N_c^2/5}$.
The numerical value of (\ref{ratio1}) for $N_c=3$ is
found as 
\begin{eqnarray}
 \left\langle \rho^3 \right\rangle_{n_\rho=0}/\rho_{\rm cl}^3 = 2.23 
\ .
\label{ratios}
\end{eqnarray}

Using (\ref{ratios}), $M_{\rm KK} = 949$ [MeV] and 
$\kappa \equiv a \lambda N_c  = 0.00745$ 
which were used in \cite{SaSu1,SaSu2}\footnote{This set of values was
obtained by fitting $f_\pi = 92.6$ [MeV] and the $\rho$ meson mass in
\cite{SaSu1}.}, 
we can evaluate the mass shift
(\ref{shiftc}) for the baryon states with $n_\rho=0$. 
We obtain 
the value of the baryon mass shift as\footnote{One could use instead the
value $M_{\rm KK} = 500$ [MeV] which was used in \cite{HSSY} for fitting
the absolute values of the baryon masses, but then it causes an
ambiguity on the choice of observable quantities one makes for getting
the value of $\lambda$. 
} 
\begin{eqnarray}
 \frac{\delta M_{n_\rho=0}}{m_\pi^2}
= 
4.11 \;{\rm [GeV^{-1}]} \ .
\label{C}
\end{eqnarray}
In particular, our result is universal for all
spin, isospin and $n_Z$.

We can compare our result (\ref{C}) for nucleons ($n_\rho=0$)
with numerical results obtained in lattice QCD simulations. 
For example, a lattice result for nucleons 
in \cite{Brommel:2008tc} shows
$\delta M_{\rm nucleon}/m_\pi^2 = 4 \times 1.02(7) \;{\rm [GeV^{-1}]}$,
and 
a summary of lattice results for nucleons in \cite{Bernard:2007zu} is
$\delta M_{\rm nucleon}/m_\pi^2  = 4 \times 0.9 \;{\rm
[GeV^{-1}]}$. Ref.~\cite{Ali Khan:2003cu} 
provides a summary of lattice QCD results in the 
year of 2003, and shows values of the coefficient as $4 \times
(0.93(5), 1.25(5),0.93(4),1.11(4))$ [GeV$^{-1}$] for four 
different fitting schemes with chiral perturbation theory. 
Rather recent results of lattice QCD
\cite{WalkerLoud:2008bp,Kadoh:2008sq} show similar values.
All of these values are consistent with ours
(\ref{C}).\footnote{However, note 
that our result is a large $N_c$ expansion with a sub-leading order term
included (for distinguishing baryon states), and also obtained under the
assumptions
which are listed in Sec.\ref{sec:2-1-2}. Since the $1/N_c$ corrections
are not so small, it is better to think of (\ref{C}) as just an order
estimate of the baryon mass shift.}

$\Delta$ excitation share the same quantum number $n_\rho=0$, thus
our value (\ref{C}) also applies to $\Delta$. A lattice result 
can be found in Ref.~\cite{Alexandrou:2008tn} as $4\times 1.20(5)$ for
$\Delta^{++}$ and $4\times 1.19(8)$ for $\Delta^+$, which are again
consistent with ours (\ref{C}).

For $n_\rho=1$, the baryon wave function is
\begin{eqnarray}
 R_1(\rho) = 
{\cal N}_1
\left(
\frac{2M_0}{\sqrt{6}} \rho^2 -1 -2\sqrt{1+N_c^2/5}
\right)
\rho^{-1+2\sqrt{1+N_c^2/5}}\exp 
\left[-\frac{M_0}{\sqrt{6}}\rho^2\right] \ .
\end{eqnarray}
In the same manner, we obtain 
\begin{eqnarray}
 \left\langle \rho^3 \right\rangle_{n_\rho=1}
&=& \left(\frac{\sqrt{6}}{2M_0}\right)^{\!\!3/2} \!\!
\frac{\displaystyle\int_0^\infty \!\!\!\! 
dt \;  \left(t -1-2\sqrt{1+N_c^2/5}\right)^2
t^{3/2+2\sqrt{1+N_c^2/5}} e^{-t}}
{\displaystyle\int_0^\infty \!\!\!\! 
dt \;  \left(t -1-2\sqrt{1+N_c^2/5}\right)^2
t^{2\sqrt{1+N_c^2/5}} e^{-t}}  
\nonumber \\
&=&
\rho_{\rm cl}^3\left(\frac{\sqrt{5}}{2N_c}\right)^{\!\!3/2} \!\!
\frac{\Gamma(b+7/2)-2b\Gamma(b+5/2)+b^2\Gamma(b+3/2)}
{\Gamma(b+2)-2b\Gamma(b+1)+b^2 \Gamma(b)}
\nonumber \\
&=&
\rho_{\rm cl}^3\left(\frac{\sqrt{5}}{2N_c}\right)^{\!\!3/2} \!\!
\frac{\Gamma(b+5/2)+\frac94 \Gamma(b+3/2)}
{\Gamma(b+1)}
\ ,
\label{ratio2}
\end{eqnarray}
which results in a relation for $N_c=3$,
\begin{eqnarray}
 \left\langle \rho^3 \right\rangle_{n_\rho=1}/\rho_{\rm cl}^3 = 4.16 \ .
\label{ratios-2}
\end{eqnarray}
Using this, the mass shift for the baryon states with $n_\rho=1$ is
obtained as 
\begin{eqnarray}
 \frac{\delta M_{n_\rho=1}}{m_\pi^2}
= 
7.65 \;{\rm [GeV^{-1}]} \ .
\label{C2}
\end{eqnarray}
This is our prediction, stating that the baryon mass shift is larger for the
$n_\rho=1$ states such as Roper excitation, while for the states with
$n_\rho=0$ such as N(1535) and $\Delta$ 
the shift is the same as that of nucleons.\footnote{ 
It was noted in \cite{HSSY} that, in the baryon spectrum of massless QCD,
there is a degeneracy in the mass spectrum for states with common value
of the sum $n_\rho+n_Z$. 
In view of our results, this degeneracy is resolved once quark masses
are turned on. Our results suggest that the mass shift of the 
$(n_\rho,n_Z)=(1,0)$ state (Roper $N(1440)$)
is larger than that of the $(0,1)$ state $N(1535)$ (which is parity odd). 
However, in reality the observed mass of the Roper is lighter than that
of $N(1535)$. This may indicate that the observed resolution of the
degeneracy in the mass spectrum is
not by the quark mass but rather would be by some higher order corrections in
$1/N_c$ and $1/\lambda$ in massless holographic QCD.
}

\section{Discussions on Meson Mass Shift}
\label{sec:3}

The shift of the baryon mass spectrum studied in the previous section
relies on an assumption that in the large 
$\lambda$ expansion the term (\ref{cma}) is the leading order term as for the effect
of the quark mass term. This term (\ref{cma}) generates a pion mass term
as seen in Sec.~\ref{sec:2-1-2}, and the assumption is equivalent to the
statement that the shift of the vector meson mass spectrum is at a
higher order in the large $\lambda$ expansion. In this section, we 
give some arguments for the assumption. The computations given here are
not sufficient to prove the assumption, but may provide some physical
intuition on how plausible the assumption is.

We give two arguments. The first one is based on the chiral perturbation
theory and its correspondence in string theory amplitudes. The second
one is a contribution coming from the D6-brane which we introduced to
obtain a finite quark mass from the worldsheet instanton (see
Sec.~\ref{sec:2-1-2}).

\subsection{Chiral perturbation and its stringy realization}

As seen in Sec.~\ref{sec:2-1-2}, the string worldsheet instanton induces
a term (\ref{mu}) of the form $c \; {\rm tr}[MU]$, which is nothing but 
the 
term for the quark mass perturbation appearing in the
chiral perturbation theory. 
It is the leading term in the derivative expansion.
In this subsection, we 
study 
other
possible corrections to the Sakai-Sugimoto effective action
(\ref{model}), from the viewpoint of chiral perturbation theory.

Sakai-Sugimoto model includes vector mesons as massive gauge fields for
hidden local symmetries, and then the covariant derivative is derived as
\begin{align}
 D_\mu U &= \partial_\mu U - iA^L_\mu(x)U + iUA_\mu^R(x) \ ,
\end{align}
and the gauge fields $A^L(x)=\cA(x,z=-\infty), A^R(x)=\cA(x,z=+\infty)$
transform as
\begin{eqnarray}
 A^L_\mu(x) &\rightarrow i g^L(x)\p_\mu g^L(x)^{-1}
+ g^L(x) A_\mu g^L(x)^{-1} \ ,
\hspace{3ex}
g^L(x) = g(x,z=-\infty) \,
\\
 A^R_\mu(x) &\rightarrow i g^R(x)\p_\mu g^R(x)^{-1}
+ g^R(x) A_\mu g^R(x)^{-1} \ ,
\hspace{3ex}
g^R(x) = g(x,z=+\infty) \ ,
\end{eqnarray}
under the gauge transformation $g(x,z)\in SU(N_f)$ on the D8-branes.

Then
we can write a higher order term which 
appears in the sense of chiral perturbation
theory as
\begin{eqnarray}
\frac{1}{f_\pi^2}
 {\rm Tr}[M U D_\mu U^\dagger D^\mu U] \ .
\label{mdudu}
\end{eqnarray}
This is, 
in the same manner as ${\rm tr}[MU]$),
invariant under the global part of the 
gauge transformation, if we simultaneously rotate the quark mass
matrix $M$, according to the philosophy of the chiral perturbation.
In 
the chiral perturbation, this is the leading
correction to the ${\rm tr}[MU]$ term in derivative 
expansion.\footnote{
The term ${\rm Tr}[MD_\mu D^\mu U]$ is at the same order, but it is
related to (\ref{mdudu}) by the equation 
$U\partial_\mu U^\dagger\partial_\mu U\sim\partial_\mu \partial^\mu U$ 
which can be obtained by a trivial identity
$\partial[U^\dagger U]=0$.
}
This term in fact generates (axial) vector meson mass terms.
It is obvious if once we expand the gauge field $\cA_\mu$ in terms of
the KK-decomposed states which are nothing but the vector meson states.

Now, our concern is the order of the coefficient of this term.
In the evaluation of the baryon mass shift in the previous section, we
simply neglected this term. As we will see, we can naively argue 
that this
term is at a higher order in the $1/\lambda$ expansion.

\setcounter{footnote}{0}

A stringy interpretation of this term can be found as follows. 
If we think of this (\ref{mdudu}) as a product of $MU$ and $D_\mu
U^\dagger D^\mu U$, the former 
comes from the worldsheet instanton amplitude, while the latter is
identical to the pion kinetic term which 
has been obtained in \cite{SaSu1,SaSu2} from the 5-dimensional YMCS 
kinetic term.
So, we can conclude that this (\ref{mdudu}) comes from a
worldsheet instanton amplitude with fluctuations of the gauge fields on
the D8-branes. More precisely, two vertex operators for the Yang-Mills
field on the D8-brane are inserted at the boundary of the worldsheet 
instanton. 

To see the order of this worldsheet instanton amplitude, 
let us briefly see a situation in the flat background. 
Worldsheet disk amplitudes in flat spacetime are of the form
\begin{eqnarray}
 A \sim \frac{1}{g_s} \int DX \exp\left[
\frac{1}{2\pi\alpha'}\int d^2\sigma \partial X \bar\partial X
-i \oint_{\rm boundary}
\hspace{-9mm} d\sigma A[X] (dX/d\sigma)
\right] \ .
\end{eqnarray}
The gauge fields here are Taylor-expanded as \cite{Fradkin:1985qd}
\begin{eqnarray}
A[X] (dX/d\sigma)= A_M[x] (dX^M/d\sigma) -\frac12  F_{MN}[x]
\widetilde{X}^N(dX^M/d\sigma)+\cdots
\label{af}
\end{eqnarray}
which is a derivative expansion, and we decomposed $X$ as 
$X=x+\widetilde{X}$ where $x$ is the zero mode of
$X[\sigma]$. The path integral can be explicitly done to this order
because the action is quadratic, and the result is
nothing but the Born-Infeld action,
\begin{eqnarray}
 A \sim \frac{1}{g_s} \sqrt{-\det(\eta + 2\pi\alpha' F)} \ .
\end{eqnarray}
It can be expanded again as
\begin{eqnarray}
 A \sim \frac{1}{g_s} \left[1 + (\alpha')^2 F^2 + \cdots\right] \ .
\label{ags}
\end{eqnarray}
When the disk spans a certain region to give the worldsheet instanton,
the zero mode $x$ dependence in the semi-classical saddle point in the
path-integral generates the worldsheet instanton factor 
$\exp [-[\rm area]/(2\pi\alpha')]$ in front of (\ref{ags}). 

Now let us try to apply this computation to our case. In our case,
we find it difficult because the background spacetime is curved, and 
also because the worldsheet boundary 
is on the D8-branes which are curved and joined. 
(See \cite{McNees:2008km} (and also \cite{Bigazzi:2004ze}) for detailed
calculations in a supersymmetric spacetime background.)
However, we can argue the order of magnitude qualitatively as follows. First, notice
that the term ``$1$'' in (\ref{ags}) in fact corresponds to our leading
order term (\ref{cma}). The reason why we have the path-ordered factor
like an Wilson loop in (\ref{cma}) is in fact the first term on the
right hand side of (\ref{af}). On the other hand, if this (\ref{ags}) is
not for a worldsheet instanton but a standard disk amplitude ({\it i.e.}
without winding some D-branes), the $F^2$ term is nothing but 
$D_\mu U^\dagger D^\mu U$, as has been found in \cite{SaSu1}.
Combining these two facts, we come to the conclusion that,
compared to the term $MU$, the term 
$MUD_\mu U^\dagger D^\mu U$ is suppressed by $(\alpha')^2$.

Since we are in the warped spacetime, this $\alpha'$ should be replaced
by an ``effective $\alpha'$'' which is given by the tension of a string 
at at the tip ($z=0$) of the geometry.\footnote{Although the gauge
fields on the D8-branes are not only at the tip, 
the wave functions of the vector mesons on the D8-branes is localized at
the tip. So we just assume that the effective worldsheet propagator in this
background geometry may be approximated by the propagator there.} This
is the QCD string tension,   
\begin{eqnarray}
 \frac{1}{2\pi \alpha'_{\rm eff}} = \frac{1}{9\pi}\lambda M_{\rm KK}^2.
\end{eqnarray}
In the Sakai-Sugimoto model, higher derivative corrections in string theory
appears with this effective $\alpha'$ (see for example discussions in 
\cite{Parthasarathy:2007tc}).  
This argument shows that the term of our concern is of the form
\begin{eqnarray}
c\; {\rm tr}
\left[MU \left(1 + {\cal O}\left(\lambda^{-2}\right)D_\mu U^\dagger D^\mu U
\right)\right] \ , 
\end{eqnarray}
corresponding to (\ref{ags}).
In the large $\lambda$ expansion which the holographic QCD employs, 
the term ${\rm tr}[MUD_\mu U^\dagger D^\mu U]$ is found to be suppressed
by $\lambda^{-2}$. 

This means that (axial) vector meson masses are not so sensitive to the
quark masses, compared to the pion mass shift.

\subsection{Effects of D6-branes}

In \cite{Hashimoto:2008sr}, to regularize the area of the worldsheet
instanton, D6-branes ending on the flavor D8-branes were introduced. 
This additional ingredient may affect the meson spectra, and in this
subsection we discuss the possible effects of the D6-branes. 

Note that if one uses a procedure of renormalization of the worldsheet
instanton proposed in \cite{Aharony:2008an}, there is no need for the 
D6-branes, so
nothing is worried about concerning the content of this subsection.
On the other hand, the D6-branes in \cite{Hashimoto:2008sr} can be
thought of as a physical cut-off which not only serves 
as a physical renormalization point but also 
provides
modes on the D6-branes.
Therefore, schematically, the effective lagrangian of the meson sector
in total should be 
\begin{eqnarray}
 S_{\rm D8} + S_{\rm D6} + S_{\rm inst}.
\end{eqnarray}
In the previous subsection we considered $S_{\rm inst}$, 
and here let us show that the term 
$S_{\rm D6}$ is sub-leading in $1/\lambda$ expansion.
This $S_{\rm D6}$ is nothing but the D6-brane effective action,
which is 
\begin{eqnarray}
 S_{\rm D6} = 
- {\cal T}_{\rm D6} \int d^7x \;
e^{-\phi}
\sqrt{-\det
(g_{MN} + 2\pi\alpha' F_{MN})} \ .
\label{D6ac}
\end{eqnarray}
Here the tension of the D6-brane is given by 
${\cal T}_{\rm D6} = 1/(2\pi)^6 l_s^7 g_s$, 
and the dilaton in the background
is $e^{-\phi} = (R/U)^{3/4}$, with the integral 
$d^7x = d^4x d\tau d\Omega_2$. Note that the D6-brane is along the 
compactified $S^1$ direction parameterized by $\tau$ which has a period
$2\pi/M_{\rm KK}$. 

In the following, we assume that the D6-brane connects the D8 and the
anti D8-brane straightly without bending. It is expected that the shape
of D6 is not straight in $\tau$ but bents, as in the case of the curved
shape of the D8-brane. However, for simplicity in this paper we assume
the straight shape of the D6-brane. 

In  addition, in the following computation, we assume that the D6-brane
wraps the largest $S^2$ in the $S^4$. As studied in an appendix of 
\cite{Hashimoto:2008sr}, the shape of the D6-brane in the $U$-$S^4$
space transverse to the $N_c$ D4-branes is non-trivial. However, since our
purpose here is to estimate the order of magnitude of the D6-brane effect, we may
take the simple D6-brane shape.

Expanding the D6-brane action (\ref{D6ac}) to quadratic order in the
field strength $F$ and substituting the background metric and dilaton, we
obtain 
\begin{eqnarray}
  S_{\rm D6} =
\frac{-1}{(2\pi)^6 l_s^7 g_s}
\int d^4x d\tau d\Omega_2 \;
R^{3/2}(U_{\rm D6})^{1/2}(2\pi\alpha')^2
\left(
\frac14 F_{\mu\nu}^2 + \frac12 f(U_{\rm D6})F_{\tau \mu}^2
\right) 
\end{eqnarray}
where $f(U)=1-U_{\rm KK}^3/U^3$, and $U_{\rm D6}$  is the location of
the D6-brane in the radial direction of the background.
Note that the D6-brane is assumed to be 
straight and wrap the $S^2$, so it is specified by just the coordinate 
$U=U_{\rm D6}$.
The action depends only on the dimensionless ratio 
$U_{\rm D6}/U_{\rm KK}$. Introducing a new variable $u \equiv U/U_{\rm KK}$,
we find that the action of the D6-brane, located at $u_{\rm D6}= U_{\rm D6}/U_{\rm KK}$, is given by
\begin{eqnarray}
   S_{\rm D6} =
\frac{-1}{(2\pi)^6 l_s^7 g_s}
\int d^4x d\tau d\Omega_2 \;
R^{3/2}(U_{\rm KK})^{1/2}(2\pi\alpha')^2
u_{\rm D6}^{1/2}\left(
\frac14 F_{\mu\nu}^2 + \frac12 (1-u_{\rm D6}^{-3})F_{\tau \mu}^2
\right) \ . \nonumber
\end{eqnarray}
Let us substitute the dictionary \cite{SaSu1} for the correspondence
between the parameters of the supergravity background 
and the QCD parameters,  
$R=(g_{\rm YM}^2 N_c l_s^2/2M_{\rm KK})^{1/3}$, 
$U_{\rm KK}=(2/9) g_{\rm YM}^2 N_c M_{\rm KK} l_s^2$ and $g_s= g_{\rm
YM}^2/2\pi M_{\rm KK} l_s$. 
Then we get
\begin{eqnarray}
 S_{\rm D6}= -\frac{1}{6\pi^2}N_c M_{\rm KK}
\int d^4x d\tau 
\; u_{\rm D6}^{1/2}
\left(
\frac14 F_{\mu\nu}^2 + \frac12 (1-u_{\rm D6}^{-3}) F_{\tau \mu}^2
\right) \ .
\label{acD6eff}
\end{eqnarray}
Note that all the $l_s$ dependence has gone.

The mode expansion of this action for eigen functions of $\tau$ gives a
spectrum of particle modes living on the D6-brane. This is similar to
the analysis of flavor D-brane fluctuations. On the other hand, since the
D6-brane ends on the D8-brane, the wave functions of the modes on the
D8-branes (which are mesons) can soak into the D6-branes, thus have
overlap with the D6-brane eigenmodes. We need to look at this
interaction and diagonalize the quadratic action to see the
possible effects on the meson mass shift.

First, let us consider the modes living on the D6-brane. 
Fourier-decomposing the gauge fields along the $\tau$ direction 
gives rise to a mass spectrum with a spacing $M_{\rm KK}$.
The eigen functions are trigonometric functions of $M_{\rm KK}\tau$,
so the integral $\int d\tau$ in (\ref{acD6eff}) cancels the factor
$M_{\rm KK}$ in front of the action (\ref{acD6eff}). This means that
properly normalized fluctuations $C_\mu(x)$ with canonical kinetic terms
is \begin{eqnarray}
 A_\mu \sim \sum_k N_c^{-1/2} \cos(k M_{\rm KK} \tau) \; C_\mu^{(k)} (x)
\label{cx}
\end{eqnarray}
up to a numerical coefficient.
The mixing can be evaluated later with this proper normalization.

Next, we look at the meson modes (on the D8-branes) which soak into the
D6-branes. We have two kinds of mesons: vector mesons whose eigen
functions $\psi_{2n}(z)$
are even in $z$, while axial vector mesons whose eigen
functions $\psi_{2n-1}(z)$
are odd in $z$. Since the D6-branes connect two points on the
D8-branes, $z=z_m$ and $z=-z_m$, so depending on the parity of the eigen
functions in $z$, the soaking into the D6-brane would be different. For
the vector mesons, the most natural soaking would be just a constant
mode on the D6-brane, while for the axial vector mesons, the natural one
which costs smallest energy would be $\cos (M_{\rm KK} \tau)$ on the
D6-brane. Once we substitute these soaking functions to (\ref{acD6eff}), 
we obtain
\begin{eqnarray}
&& S_{\rm D6}= 
\frac{-N_c}{6\pi}
\int d^4x \; u_{\rm D6}^{1/2}
\left[
\frac14 (F_{\mu\nu}^{\rm (even)}(z=z_m))^2
\right.
\nonumber \\
&&
\left.
\hspace{20mm}
+
\frac18 (F_{\mu\nu}^{\rm (odd)}(z=z_m))^2
+
\frac12 M_{\rm KK}^2 (1-u_{\rm D6}^{-3}) (A_{\mu}^{\rm (odd)}(z=z_m))^2
\right]
\label{evenodd}
\end{eqnarray}
Here, we divided the wave functions into the odd part and the even part,
\begin{eqnarray}
&& A_\mu (x,z)=A_\mu^{\rm (odd)} +  A_\mu^{\rm (even)},
\\
&& 
A_\mu^{\rm (odd)} = \sum_{n=1}^{\infty}
B_\mu^{(2n)}(x) \psi_{2n}(z), 
\quad
A_\mu^{\rm (even)} = \sum_{n=1}^{\infty}
B_\mu^{(2n-1)}(x) \psi_{2n-1}(z).
\end{eqnarray}

Note that to obtain the expression (\ref{evenodd}) we identified 
the gauge fields on the D6-brane with that of the D8-branes. 
The reason why there is no difference in the normalizations 
is as follows. Let us consider a D$p$-brane ending on a D$(p+3)$-brane.
The D$p$-brane can be thought of as a spike solution of a BPS Dirac
monopole on the D$(p+3)$-brane 
\cite{Callan:1997kz}. The $A_\mu$ ($\mu=0,\cdots,p$) 
fluctuations on the D$(p+3)$-brane can soak onto the D$p$-brane.
The DBI action of the D$(p+3)$-brane is, if we turn off the other 
fluctuations,
\begin{eqnarray}
 S_{p+3} = {\cal T}_{{\rm D}(p+3)}\int d^{p+4}x 
\sqrt{\det (\eta_{\mu\nu} + 2\pi\alpha' F_{\mu\nu})}
\sqrt{\det (\delta_{ij} + \partial_i X^{(c)} \partial_j X^{(c)} 
+ 2\pi \alpha' F_{ij}^{(c)})}
\nonumber 
\end{eqnarray}
where $X^{(c)}$ and $A^{(c)}_i$ are classical solution for the spike,
and $i,j=p+1,p+2,p+3$.
Explicit substitution of this classical solution gives
\begin{eqnarray}
 {\cal T}_{{\rm D}(p+3)}\int dx^{p+1} dx^{p+2} dx^{p+3} 
\sqrt{\det (\delta_{ij} + \partial_i X^{(c)} \partial_j X^{(c)} 
+ 2\pi \alpha' F_{ij}^{(c)})}
={\cal T}_{{\rm D}p} \ ,
\end{eqnarray}
therefore we obtain
\begin{eqnarray}
 S_{p+3} = {\cal T}_{{\rm D}p}\int d^{p+1}x 
\sqrt{\det (\eta_{\mu\nu} + 2\pi\alpha' F_{\mu\nu})}
\end{eqnarray}
which is the standard D$p$-brane action. This shows that the gauge field
on the D8-brane is in fact a gauge field on the D6-brane, without
changing its normalization.

The normalization of the eigen functions $\psi_n(z)$ is defined 
in \cite{SaSu1} as 
\begin{eqnarray}
\frac{g_{\rm YM}^2 N_c^2}{216 \pi^3}
\int dz K^{-1/3} \psi_n(z)\psi_m(z) = \delta_{nm}.
\end{eqnarray}
So, defining the normalized eigen function
\begin{eqnarray}
\tilde{\psi}_n(z)\equiv \sqrt{\frac{g_{\rm YM}^2 N_c^2}{216 \pi^3}}
\psi_n(z),
\label{redefpsi}
\end{eqnarray}
we have the standard normalization
$\int dz K^{-1/3} \tilde{\psi}_n(z)
\tilde{\psi}_m(z) = \delta_{nm}$.
Substituting this normalization to the D6-brane action, we obtain
\begin{eqnarray}
&& S_{\rm D6}= 
\frac{-9\pi^2 u_{\rm D6}^{1/2}}{\lambda}
\sum_{n,m=1}^\infty \tilde{\psi}_{2n-1}(z_m)
\tilde{\psi}_{2m-1}(z_m)
\int d^4x \; 
\left[
\frac14 F_{\mu\nu}^{(2n-1)} F_{\mu\nu}^{(2m-1)}
\right]
\nonumber \\
&&
+
\frac{-9\pi^2 u_{\rm D6}^{1/2}}{\lambda}
\sum_{n,m=1}^\infty \tilde{\psi}_{2n}(z_m)
\tilde{\psi}_{2m}(z_m)
\int d^4x \; 
\left[
\frac18 F_{\mu\nu}^{(2n)}F_{\mu\nu}^{(2m)}
+
\frac12 M_{\rm KK}^2 (1-u_{\rm D6}^{-3}) 
B_{\mu}^{(2n)}B_{\mu}^{(2m)}
\right].
\nonumber 
\\
&&
\label{BB}
\end{eqnarray}
Here $F$ are the field strengths of the (axial) vector mesons
$B_\mu(x)$. 

The information we need to extract from 
the last expression (\ref{BB}) is its
overall factor $1/\lambda$. The action (\ref{BB}) is supposed to be
added to the D8-brane action (which is the meson action of
\cite{SaSu1}), so we find a $1/\lambda$ correction due to the D6-brane. 

There is an overlap between these soaking modes of the mesons and the
modes on the D6-brane (\ref{cx}). Since the former has a normalization
factor $\lambda^{-1/2} N_c^{-1/2}$ as seen in (\ref{redefpsi}) while the
latter (\ref{cx}) has $N_c^{-1/2}$, substituting them to
(\ref{acD6eff}), we find the overlap giving a
quadratic interaction of order ${\lambda}^{-1/2}$.

Therefore, in total, we obtain the
kinetic terms for
the modes on the
D8-branes (mesons) and the modes on the D6-brane schematically written
as 
\begin{eqnarray}
 S \sim \int \! d^4x \ (F^{\rm D8}, F^{\rm D6})
\left(
\begin{array}{cc}
1 + {\cal O}(\lambda^{-1})  & \;\;{\cal O}(\lambda^{-1/2})\\
{\cal O}(\lambda^{-1/2}) & 1
\end{array}
\right)
\left(
\begin{array}{c}
F^{\rm D8} \\
F^{\rm D6}
\end{array}
\right) \ .
\end{eqnarray}
Diagonalizing this matrix, we obtain a shift to the vector meson
mass which is of order $
{\cal O}(\lambda^{-1/2})$.
As advertised, it is a sub-leading order term, compared to the
original kinetic terms in the D8-brane action for the vector mesons.
Furthermore, this shift is further suppressed by the 
wave function factor $\tilde{\psi}(z_m)$.

\section{Summary}

In this paper, we computed the shift of the
baryon mass spectrum by varying the quark mass, from the chiral limit,
in the Sakai-Sugimoto model \cite{SaSu1, SaSu2}
of holographic QCD. To introduce the quark mass to the model,
we employed the worldsheet instanton approach 
\cite{Aharony:2008an,Hashimoto:2008sr}. We combined it with 
the solitonic description of the baryon used in \cite{HSSY}, 
and we obtained analytically the mass shift of the baryons, which is linear
in the pion mass squared. The slope depends on the species of
the baryons. 

The slope computed for nucleons (whose value 
is also shared with $\Delta$),
with inputs chosen to be 
$f_\pi$ and 
$m_\rho$ as in \cite{SaSu1, SaSu2}, roughly agrees with lattice data for
nucleons and $\Delta$.
We find that the slope is shared also with $N(1535)$, while the
slope for the Roper excitation $N(1440)$ is found to be larger. These
slopes for excited baryons are our predictions, and it would be quite
interesting if they can be confirmed by lattice computations in the
future. Although we work in large $N_c$ QCD with the holographic
approach, we hope that the tendency of the difference in the slope 
may be the same for QCD.

In Sec.~3, we gave various discussions on the vector meson mass
shift, but our naive arguments showed that possible effects are
subleading order in the large $\lambda$ expansion
(which justifies our computation in Sec.~2). It would be important
to clarify this issue further
and compute the vector meson mass shift more rigorously.

\acknowledgments 
K.H.~and T.H.~would like to thank Summer Institute 2008 at Chi-Tou,
Taiwan, at which a part of this work was discussed.
K.H.~likes to thank Daisuke Kadoh for explaining lattice QCD results, and
is partly supported by
the Japan Ministry of Education, Culture, Sports, Science and
Technology. The work of D.K.H.~is
supported by the Korea Research Foundation Grant funded by the Korean
Government 
(MOEHRD, Basic Research Promotion Fund) (KRF-2007-314- C00052). T.H.~is
supported by National Center for Theoretical Sicnences, Taiwan (No. NSC
97-2119-M-002-001, NSC97-2119-M-007-001) . 


\end{document}